\documentclass[reprint,prb,superscriptaddress,amsmath,amssymb,aps]{revtex4-2}

\usepackage{graphicx}
\usepackage{dcolumn}
\usepackage{bm}
\usepackage{lipsum}
\usepackage{color}
\usepackage{subfigure}
\usepackage{caption}
\usepackage[colorlinks=true,citecolor=blue,linkcolor=blue]{hyperref}

\begin{document}

\title{Seebeck and Nernst effects in topological insulator: the case of strained HgTe}
\author{Francisco J. Pe\~{n}a}\email{francisco.penar@usm.cl}
\affiliation{Departamento de F\'{i}sica, Universidad T\'{e}cnica Federico Santa Mar\'{i}a, Casilla 110-V2390123 Valpara\'{i}so, Chile}
\author{Oscar Negrete}\email{oscar.negrete@usm.cl}
\affiliation{Departamento de F\'{i}sica, Universidad T\'{e}cnica Federico Santa Mar\'{i}a, Casilla 110-V2390123 Valpara\'{i}so, Chile}
\affiliation{Centro para el Desarrollo de la Nanociencia y la Nanotecnolog\'{i}a, 8320000 Santiago, Chile}
\author{Ning Ma}\email{maning@tyut.edu.cn}
\affiliation{Department of Physics, College of Physics and Optoelectronic Engineering, and MOE Key Laboratory of Advanced
Transducers Intelligent Control System, Taiyuan University of Technology,
Taiyuan 030024, China}
\author{Patricio Vargas}\email{patricio.vargas@usm.cl}
\affiliation{Departamento de F\'{i}sica, Universidad T\'{e}cnica Federico Santa Mar\'{i}a, Casilla 110-V2390123 Valpara\'{i}so, Chile}
\affiliation{Centro para el Desarrollo de la Nanociencia y la Nanotecnolog\'{i}a, 8320000 Santiago, Chile}
\author{M. Reis} \email{marioreis@id.uff.br}
\affiliation{Institute of Physics, Fluminense Federal University, Av. Gal. Milton Tavares de Souza s/n, 24210-346, Niteroi-RJ, Brazil}
\author{Leandro R. F. Lima}\email{leandrolimaif@gmail.com}
\affiliation{Departamento de F\'{\i}sica, Instituto de Ci\^encias Exatas, Universidade Federal Rural do Rio de Janeiro, 23897-000 Serop\'edica - RJ, Brazil}
\begin{abstract}
We theoretically study the thermoelectric transport properties of strained HgTe in the topological insulator phase.
We developed a model for the system using a Dirac Hamiltonian including the effect of strain induced by the interface between HgTe and the CdTe substrate. The conductivity tensor was explored assuming the electrons are scattered by charge impurities, while the thermopower tensor was addressed using the Mott relation. Seebeck and Nernst responses exhibit remarkable enhancements in comparison with other two-dimensional Dirac materials, such as graphene, germanane, prosphorene and stanene. The intensity of these termoeletric responses, their dependencies with the external perpendicular magnetic field and temperature are also addressed.
\end{abstract}


\maketitle

\section{Introduction}

The studies of the simultaneous electric and heat currents in a conductor have been helped by the macroscopic treatment of coupled irreversible fluxes developed by Onsager \cite{macdonald2006thermoelectricity}, supporting the understanding of the physics behind the \textit{thermoelectric} (TE) effects. These materials are characterized by the ability of converting (wasted) heat into electricity, by exposing the material terminals to a temperature gradient: this is the Seebeck effect \cite{rothe2012spin,uchida2008observation,ramos2017spin}. The contrary effect is also possible, i.e., a voltage between the material terminals converted into temperature gradient, named as Peltier effect \cite{gratz2002electrical}. TE materials have been used and proposed to be used in several technological applications, including energy harvesting/generation \cite{rowe2018thermoelectrics,champier2017thermoelectric}, sensing \cite{han2017thermoelectric,feng2019flexible} and even wearables \cite{hong2019wearable}, just to name a few. 

An alternative situation in which the current flows transversely to both the applied magnetic field and the temperature gradient corresponds to the Nernst effect \cite{rothe2012spin}. This last effect has taken a high interest in the latest time because materials with a high Nernst coefficient could be used in cryogenic refrigeration \cite{freibert2001thermomagnetic}. This has prompted the search for materials that present this property, highlighting among them those observed in correlated electron systems  \cite{behnia2009nernst,bel2004giant,bel2004thermoelectricity}, conventional semi-metals \cite{behnia2007nernst}, metallic ferromagnets \cite{ramos2014anomalous,lee2004anomalous,miyasato2007crossover} as well as in Dirac/Weyl semi-metal materials \cite{yang2020giant}.

The TE effects strongly depend on the density of states of the substance and, consequently, topological insulators (TIs) in two- (2D) and three- dimensions (3D) have been attracting much attention of the community \cite{xu2017topological}. TIs present attractive features, such as gaped bulk states and, simultaneously, gapless Dirac boundary states \cite{xu2017topological}. This topological protection produces a remarkable phenomenon, as the spins are locked with the momentum, and therefore the charge carriers on the protected states do not suffer scattering due to defects \cite{xu2017topological}. Nevertheless, 3D TIs have strong defect doping and low carrier mobility in the experiments; and, consequently, the bulk conductivity obscures the surface charge transport. Thus, the search for the regime in which the bulk carriers are absent is now the central focus of the field.

In this direction, experiments carried out in strained 3D HgTe have shown that this material corresponds to a TI because the applied strain opens a gap between the light-hole and heavy-hole bands \cite{brune2011quantum}. Moreover, transport measurements in strained 3D HgTe due to CdTe substrate show that the Landau levels (LLs) remain degenerate as long as the hybridization can be neglected between the top and bottom surface states (e.g., HgTe 70 nm of thickness) \cite{zhang2011band,zhang2009topological}. Several analytical models have been proposed to explain the experimental measurements employing the model of two-Dirac cones, that adequately account for the separation of the Hall plateaus, the Shubnikov-de Haas oscillations (SdH) due to strain, and the contribution of zero energy states in the conductivity \cite{shan2010effective,brune2011quantum,buttner2011single,tahir2012quantum}. In addition, various theoretical and experimental studies have emerged, trying to figure out the contribution of TIs boundary states to thermoelectricity. We highlight those carried out in bismuth telluride (Bi$_{2}$Te$_{3}$), antimony telluride (Sb$_{2}$Te$_{3}$), bismuth selenide (Bi$_{2}$Se$_{3}$) and tin telluride (SnTe), where these materials showed excellent TE performance \cite{hsieh2009observation,hsieh2012topological,chen2009experimental,zhang2011band,zhang2009topological}.

In the present work, we study two TE effects, the Seebeck and the Nernst effects for HgTe film, strained due to the CdTe substrate under the action of a perpendicular external magnetic field over the sample. This material was modeled by a two-dimensional non-ideal Dirac quasi-particle Hamiltonian. Using the results obtained for collisional and Hall conductivity from Kubo-Greenwood formalism \cite{vasilopoulos1984,vanvliet1988,peeters1992}, we calculated this material's TE properties using the Mott relation \cite{mott1979,jonson1980}. Our results indicate that an oscillatory Seebeck is obtained due to the SdH oscillations present in the collisional conductivity and strongly modulated by the substrate-induced strain. In the case of the Nernst effect, we found a nearly zero (or zero) value around the charge neutrality point, which is a clear manifestation of a non-trivial topological transition between the top and bottom surfaces valleys. Finally, in a low-temperature regime, we found a maximum value for the Seebeck of $\sim 3.2\ k_{B}/e$ and the Nernst effect close to $1.2\ k_{B}/e$, which is in accordance with other Dirac materials.

Our paper is organized as follows: section~II shows the theoretical model for the calculation of conductivity and its implication on the TE properties of the material. Section~III displays the results and discussions for both Seebeck and Nernst effects. Finally, section~IV shows the concluding remarks for this work.

\section{Electrical conductivity}

In this section, we will present the proposed device for the measurements of the Seebeck and Nernst coefficients. Additionally, we will discuss in-depth the theoretical model employed and the approximations used to calculate the collisional and Hall conductivity. This discussion is useful for the results reported in the later sections about the thermoelectric coefficients.

\begin{figure}
\centering
\includegraphics[width=1\linewidth]{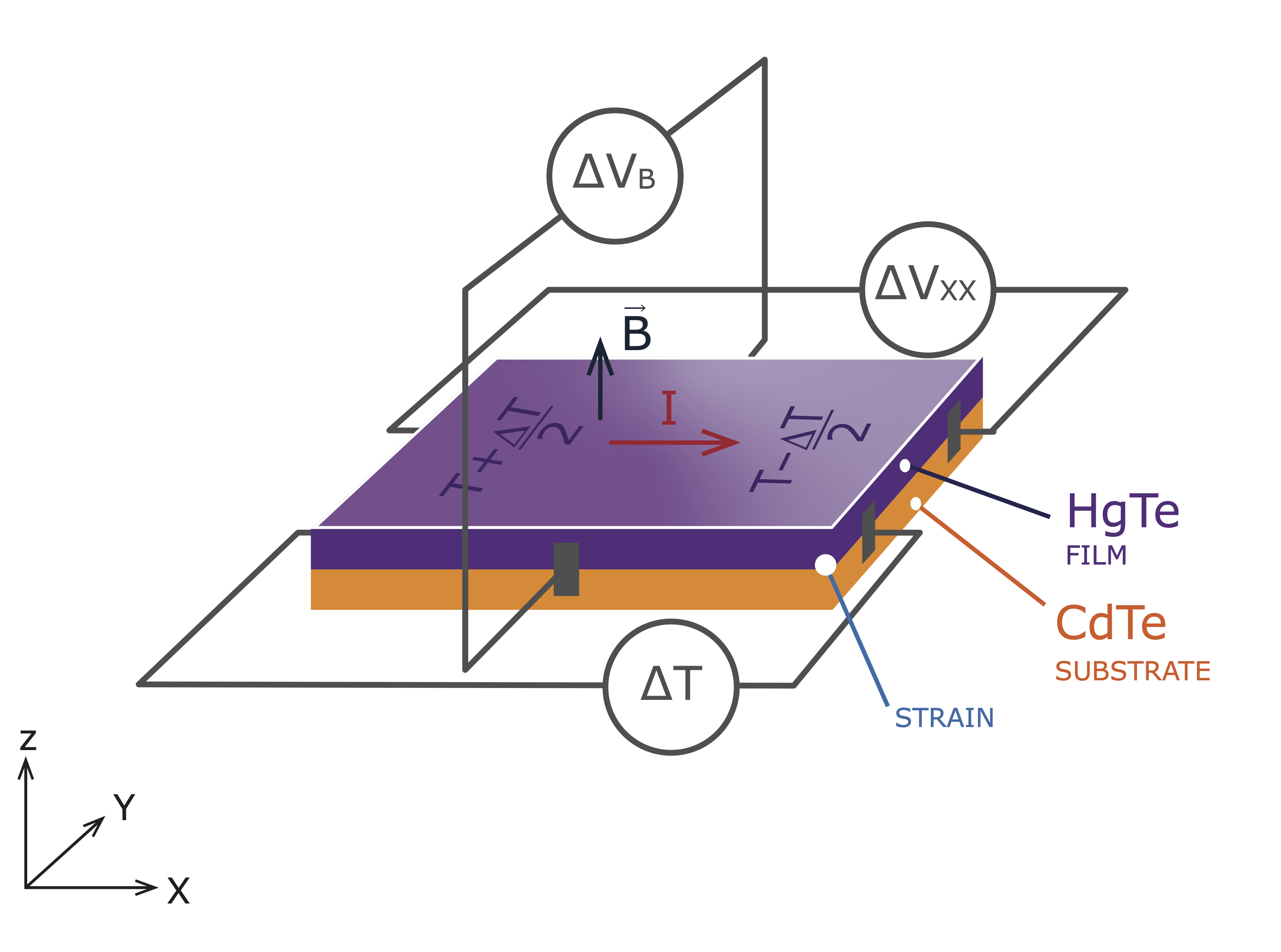}
\caption{Proposed device. The HgTe film deposited on CdTe substrate. The electronic current $I$ and the voltage $\Delta V$ are responses to the temperature gradient $\Delta T$ and the perpendicular magnetic field $\mathbf B$. See the text for further details.}
\label{device}
\end{figure}

The proposed device is composed of a thick HgTe film, strained due to the CdTe substrate. Figure~\ref{device} shows the device designed. The film has two surfaces, the one at the top, facing vacuum, and the other at the bottom, facing CdTe substrate. The thickness of the film is such that there is no hybridization between the surface states. The temperature gradient $\Delta T$, applied between the ends of the system, originally at a temperature $T$, drives a longitudinal electronic current $I$ (equivalent to a bias). The presence of a perpendicular magnetic field $\mathbf B$ induces an electronic current (bias) that is transverse to both the magnetic field and the temperature gradient. Those longitudinal and the transverse induced biases (with respect to the temperature gradient) are known as Seebeck and Nernst effects, respectively. In what follows we quantify these effects by calculating the proper coefficients related to the electronic conductivity of the system. 

The TE effects are related with the thermopower $S$, and can be calculated from the Mott relation \cite{mott1969TE} as
\begin{align}
\label{thermopower}
S=-\frac{\pi ^{2}k_{B}^{2}T}{3e}\frac{\partial \ln \sigma \left( \mu \right)}{\partial \mu },
\end{align}
where $\mu$ is the chemical potential and $\sigma$ is the 2D conductivity tensor. We model our device and calculate the 2D conductivity tensor using the advanced models presented in Refs.~\cite{tahir2012quantum,tahir2013quantum,tahir2013valley,ning2016TI,tahir2016}, as described below.
The 2D conductivity tensor of the device presented in Fig.~\ref{device} can be calculated using a two Dirac cones model  as presented by M. Tahir and U. Schwingenschl\"{o}gl in Refs.~\cite{tahir2012quantum,tahir2013quantum,tahir2013valley}.N. Ma et al. modified this theory \cite{ning2016TI} to further explain features of the transport measurements, such as the well separated surface-dependence Hall plateaus and SdH oscillations due to the strain, the contribution of zero energy states, and specially, the strain induced changes in the zero-mode conductivity at different surfaces. 

However, the correction made in Ref. \cite{ning2016TI}  is valid for small strain values. When the strain is large, it presents irregularities in the collisional conductivity calculations such as a high asymmetry in the peaks associated with the first Landau levels.  This is why we take the path proposed by M. Tahir and U. Schwingenschl\"{o}gl  from Ref. \cite{tahir2012quantum}, with the correction made by the same authors years later in Ref. \cite{tahir2016}. This will be the approach that will be detailed below.

We model the propagating surface states lying on the $x-y$ plane
using a two Dirac cones model \cite{Molenkamp2011TI} in the form of a 2D non-ideal Dirac quasi-particle Hamiltonian, given by:
\begin{align}
\label{hamiltonian11}
H=\tau_{z}\otimes\upsilon _{F}\left( \mathbf{\sigma }_{x}\mathbf{\pi }%
_{y}-\mathbf{\sigma }_{y}\mathbf{\pi }_{x}\right) +\tau_{z}\otimes\mathbf{%
I}\Delta .
\end{align}
The first term arises from the spin-orbit coupling (SOC) that induces a unique spin-momentum locking, which means that the electrons with opposite spins travel in opposite directions \cite{tokura2019magnetic}. This is essential for modeling topologically
nontrivial insulators. 
Here the parameter $\tau_{z} =1$ and $\tau_{z} = -1$ represent the surfaces facing vacuum and CdTe substrate, respectively; $\upsilon _{F}=4.0\times 10^{5}$ m/s is the Fermi velocity of the surface states in HgTe, which is smaller than the Fermi velocity in graphene ($1.0\times 10^{6}$ m/s) \cite{castroneto2009}, $\sigma_x$ and $\sigma_y$ are the $x$ and $y$ matrix components of the Pauli vector, and $\boldsymbol\pi =\mathbf p+e 
\mathbf{A}$ is the canonical momentum, where $\mathbf{A}$ the vector potential in the Landau gauge $\mathbf{A}=\left(0,Bx,0\right)$, which leads to to an external magnetic field $\mathbf{B=}\left( 0,0,B\right)$. 
The second term results from the strain between the HgTe and CdTe films.
The strain energy $\Delta$ acts on both surface states with opposite signs. 
Here, $\mathbf{I}$ is the identity matrix.

The eigenvalues of the Hamiltonian are given by
\begin{equation}
\label{energyspectrum}
E_{n,\lambda }^{\tau_{z}}=
\begin{cases}
\lambda \hbar \omega _{c}\sqrt{2 n}+\tau_{z} \Delta, & n > 0\\
    \tau_{z}\Delta,              & n=0
\end{cases}
\end{equation}
where $\lambda=+1$ holds for electron and $\lambda =-1$ for hole bands. The integer $n$ $(n=0,1,2,...)$ represents both the Landau Levels (LLs) and the number of conducting channels of the 2D edge quantum states. For 3D topological insulator, $n$ is actually the topological invariant, named as the ``first Chern number'' in the momentum state space. The cyclotron frequency is given by $\omega _{c}=\upsilon _{F}/\ell _{c}$, where $\ell _{c}=\sqrt{\hbar /eB}$ is the magnetic length. 

\begin{figure}
\begin{center}
\includegraphics[width=9cm]{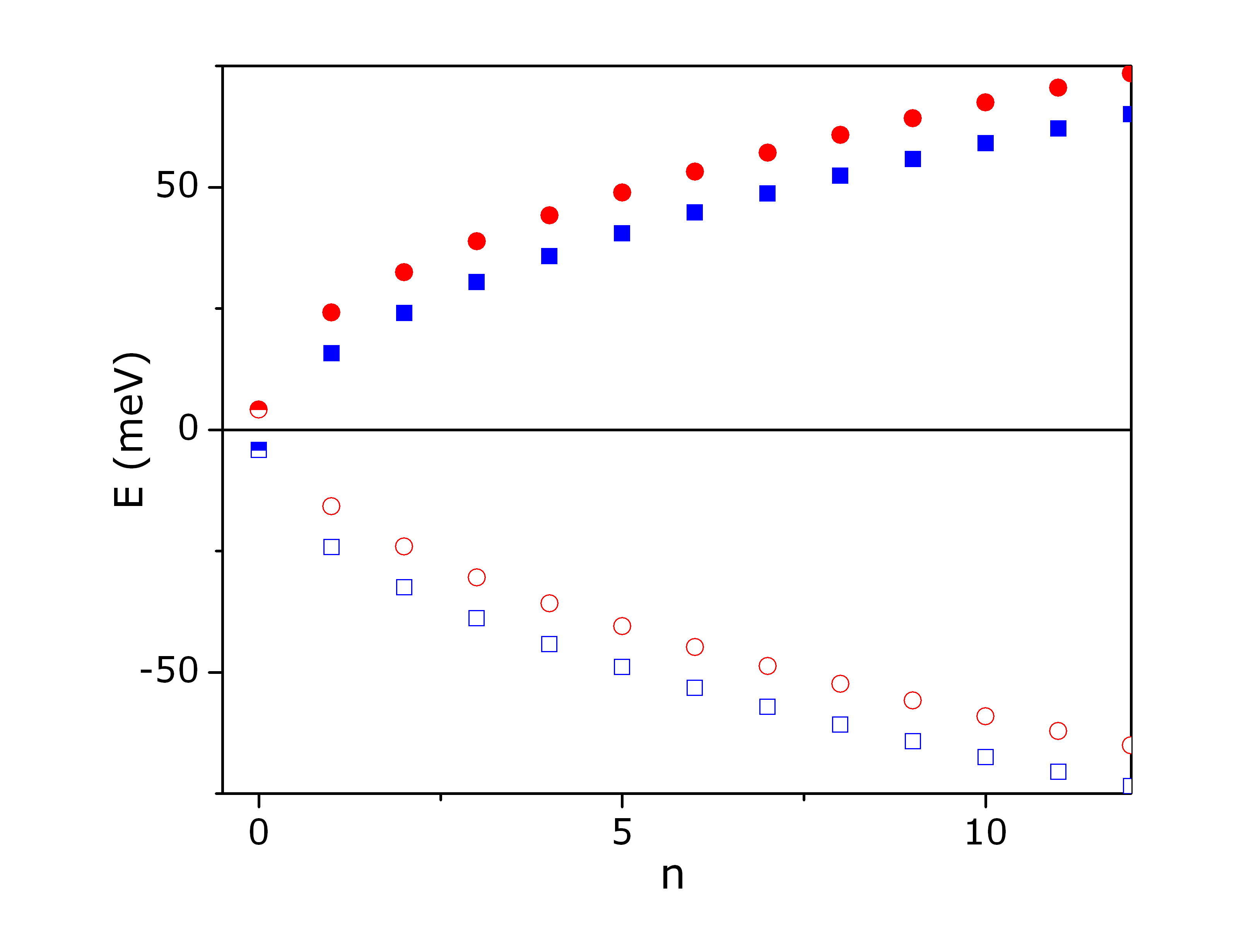}
\caption{Energy spectrum as a function of quantum number $n$ for electrons and holes from bottom and top surfaces.  Here $B=2$ T and $\Delta=4.2$ meV. }
\label{energyspectra}
\end{center}
\end{figure}

Figure~\ref{energyspectra} presents the energy spectra of Eq.~(\ref{energyspectrum}) as a function of the LL index $n$, for both electrons and holes ($\lambda=\pm1$) and top and bottom surfaces ($\tau_{z}=\pm1$).
We notice that the zero mode at the same surface (top/bottom) is not gapped, indicating the present strain could not give rise to the topological phase transition. 
However, the strain opens up a gap $2\Delta$ between the $0$-th levels from opposite surfaces.
This gap is fundamental for the study of collisional conductivity close to charge neutrality point (CNP) and will be discussed later.
Notice that since the strain shifts the two surfaces' Dirac cones toward opposite energy directions (positive/negative), it breaks both inversion symmetry and valley degeneracy.

The 2D conductivity tensor is defined as
\begin{align}
\label{sigma}
\sigma \left( \mu \right) =\left(
\begin{array}{cc}
\sigma _{xx} & \sigma _{xy} \\
\sigma _{yx} & \sigma _{yy}%
\end{array}%
\right)
\end{align}
where $\sigma _{yy}=\sigma _{xx}$ and $\sigma _{yx}=-\sigma _{xy}$. 
Using the model Hamiltonian in Eq.~(\ref{hamiltonian11}), we follow Refs.~\cite{tahir2012quantum,tahir2013quantum,tahir2013valley,ning2016TI,tahir2016} to calculate the collisional conductivity $\sigma _{xx}$ and the Hall conductivity $\sigma_{xy}$ as
\begin{widetext}
\begin{align}
\label{sigmaxx}
\sigma _{xx} &=\frac{e^{2}}{h}\frac{  N_{I} e^{4}}{8\pi^{2}\epsilon^{2}l_{c}^{2}k_{s}^{2}\hbar \omega_{c}k_BT}\sum\limits_{n,\tau_{z},\lambda}\left[\left(2n+1\right) \cos^{4}\left(\frac{\theta}{2}\right) + (2n-1)\sin^{4}\left(\frac{\theta}{2}\right)-2n
\cos^{2}\left(\frac{\theta}{2}\right)\sin^{2}\left(\frac{\theta}{2}\right)\right]{\text{sech}^{2}\left(\xi^{\tau_z}_{n,\lambda}\right)}, \\
\label{sigmaxy}
\sigma _{xy} &=\frac{e^{2}}{h}\sum\limits_{n,\tau_{z},\lambda}\sin^{2}\left(\theta\right)\left( n+\frac{1}{2}%
\right) \left[\tanh\left(\xi_{n+1,\lambda}^{\tau_{z}}\right) -\tanh\left(\xi_{n,\lambda}^{\tau_{z}}\right) \right],
\end{align}
\end{widetext}
where $\xi^{\tau_{z}}_{n,\lambda} = (E_{n,\lambda}^{\tau_{z}}-\mu)/2k_{B}T$, $\theta =\tan^{-1}\left(\frac{\sqrt{n}\hbar \omega}{\Delta}\right)$, $N_I$ stands for the impurity density, $\epsilon=\epsilon_r\epsilon_0$ is the dielectric constant of the material, $\epsilon_r$ is the relative permittivity, $\epsilon_0$ is the vacuum permittivity and $k_s$ is the screening vector \cite{tahir2012quantum}. Unless stated otherwise, we use $\epsilon_{0}=8.854\times 10^{-12}$, $\epsilon_{r}=4$, $N_{I}=1\times 10^{15}$ m$^{-2}$ and $k_{s}=1\times 10^{7}$ m$^{-1}$ throughout the paper \cite{tahir2012quantum}.

It is important to mention that we included the correction made in Ref.~\cite{tahir2016}, as pointed out by Ref.~\cite{ning2016TI}. 
Although this correction term does not significantly change the general behavior of the main results, it must be added since the thermopower intensity is directly related to it, due to Eq.~(\ref{thermopower}). 

\begin{figure}
\subfigure[]{\includegraphics[width=8cm]{{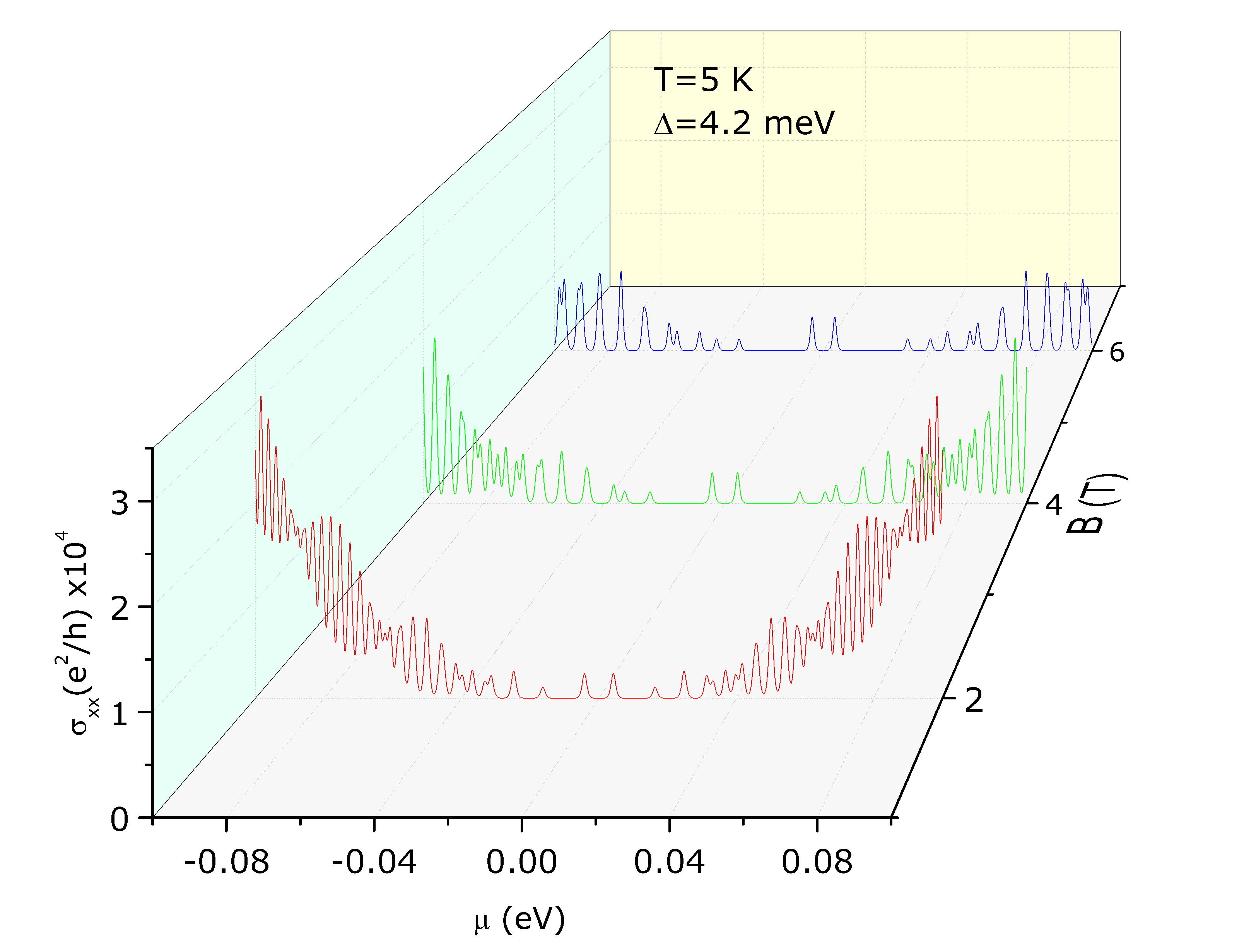}}}\hspace{0.5cm}
\subfigure[]{\includegraphics[width=8cm]{{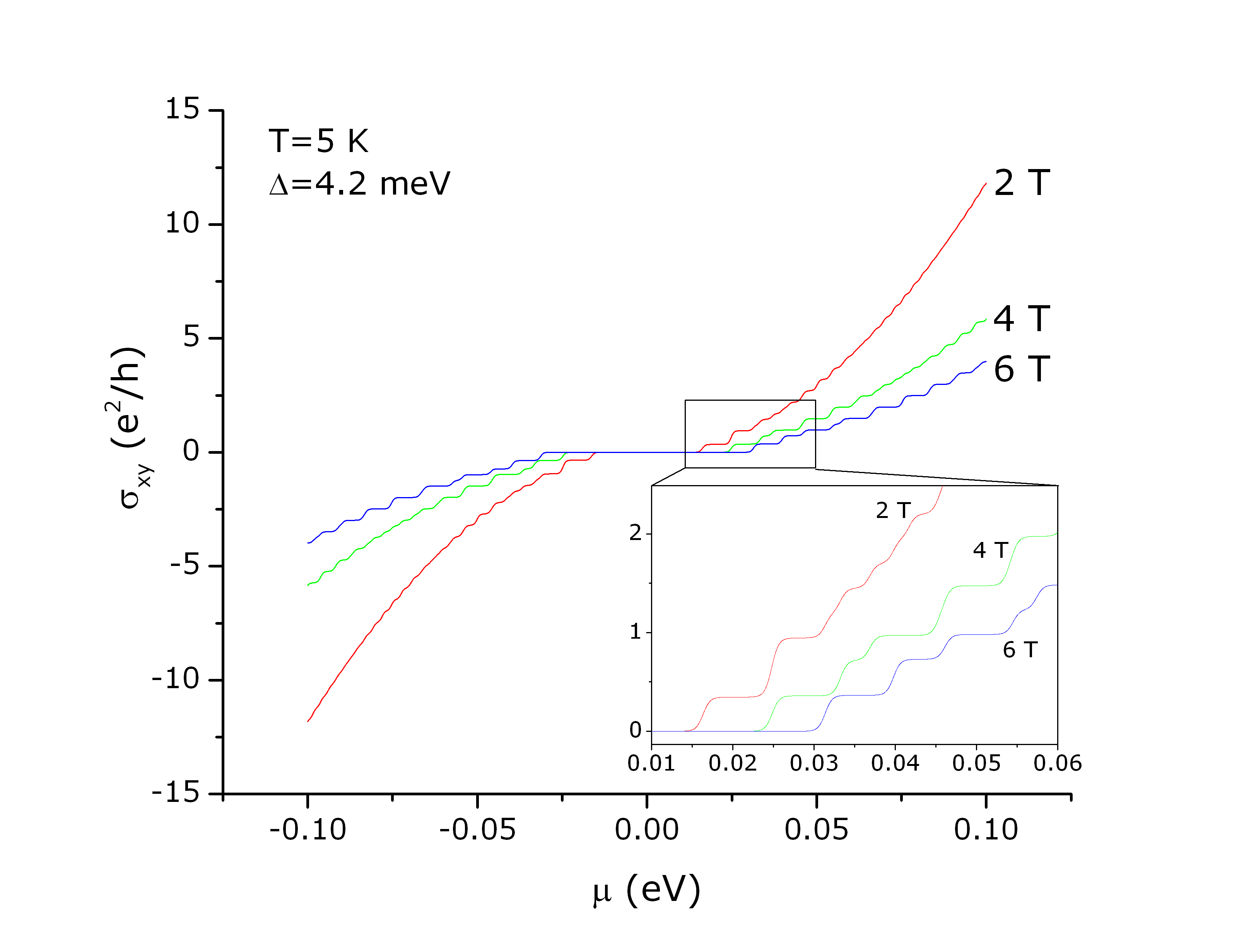}}}
\caption{(a) Collisional $\sigma_{xx}$ and (b) Hall $\sigma_{xy}$ conductivities as functions of chemical potential $\mu$ for $B=2$ T, $4$ T and $6$ T.  Here we used $\Delta=4.2$ meV and $T=5$ K. }
\label{figsigmamub}
\end{figure}

Figure~\ref{figsigmamub} shows the collisional conductivity $\sigma_{xx}$ and the Hall conductivity $\sigma_{xy}$ as functions of the chemical potential $\mu$ for different values of the magnetic field $B$. The oscillations of $\sigma_{xx}$ in Fig.~\ref{figsigmamub}(a) reflect the periodicity of the SdH oscillations. The SdH peaks occur when the chemical potential coincides with the LLs. The intervals between the peaks are unequal since the LLs are unequally spaced according to Eq.~(\ref{energyspectrum}). The SdH amplitude increases with the increasing of the chemical potential due to the larger scattering rate of LLs with higher index/filling factor at a given magnetic field. 
Moreover, the overlap between two consecutive peaks increases with the chemical potential as the energy separation between higher energy LL's become smaller as $n$ increases.
Due to strain, the single peak splits in two, with a gap opening at the CNP and a well-resolved beating pattern of SdH oscillations appearing away from the CNP. 
These beating pattern indicate that the SdH oscillations of different surfaces are out-of-phase.
The broken valley degeneracy in the Landau spectra of the two surfaces, caused by the applied strain, leads to different electrostatic environments.
At small chemical potential values, the LL's are far apart having very little overlap. 
As the chemical potential increases, the LL's become closer and overlap with each other. 
The matching of two LL's from different surfaces creates an enhanced peak corresponding to a maximum of the beating pattern.
The nodes of the beating pattern follow the theoretical approximation for zero temperature given by $\mu_l = \hbar^2\omega_c^2l/(8\Delta)$, where $l$ is an odd integer \cite{tahir2012quantum}.
A further increase of the magnetic field weakens the beating pattern until it disappears as the LL's separation $\hbar\omega_c$ is proportional to $\sqrt{B}$. 

Figure~\ref{figsigmamub}(b) shows the Hall conductivity $\sigma_{yx}$ for different values of $B$, as a function of $\mu$. We see that the Hall conductivity is strictly quantized due to the quantized LLs. It increases one by one in the unit of $e^2/h$ with the increasing of $\mu$ since the LLs are filled one by one. Therefore, we observe the integer Hall plateaus at $0, \pm 1, \pm 2, \pm 3, \pm 4,..., $ in Hall conductivities. For a given magnetic field, the width of the plateaus is unequal since the LL spacing of two adjacent LLs are unequal, according to Eq.~(\ref{energyspectrum}).  As shown and discussed in reference \cite{ning2016TI}, under strain, some extra Hall plateaus arise. These new plateaus coincide with the sharp peaks of $\sigma_{xx}$, and have their origin in the strain. In association with the strong magnetic field, it breaks two surfaces inversions symmetry and thus removes the valley double degeneracy in their Landau spectra. Thus, the density of states forms different Landau ladders for different surfaces. These extra quantum plateaus take values at even filling factor $0, \pm 2, \pm 4, \pm 6, \pm 8,..., $ etc, which have been already confirmed by the experiments \cite{Molenkamp2011TI}.


In the limit of zero temperature and zero strain energy, the Hall conductivity reads as:
\begin{equation}
    \sigma_{xy}=\frac{2e^2}{h}\left(n+\frac{1}{2}\right).
\end{equation}
Above, the pre-factor 2 results from the surface degeneracy. Note the Hall plateaus appear at the filling factor $\pm 1, \pm  3, \pm 5, ..., $ in agreement with transport experiments \cite{Zhangsc2011TI}. 

\begin{figure}
\begin{center}
\subfigure[]{\includegraphics[width=8cm]{{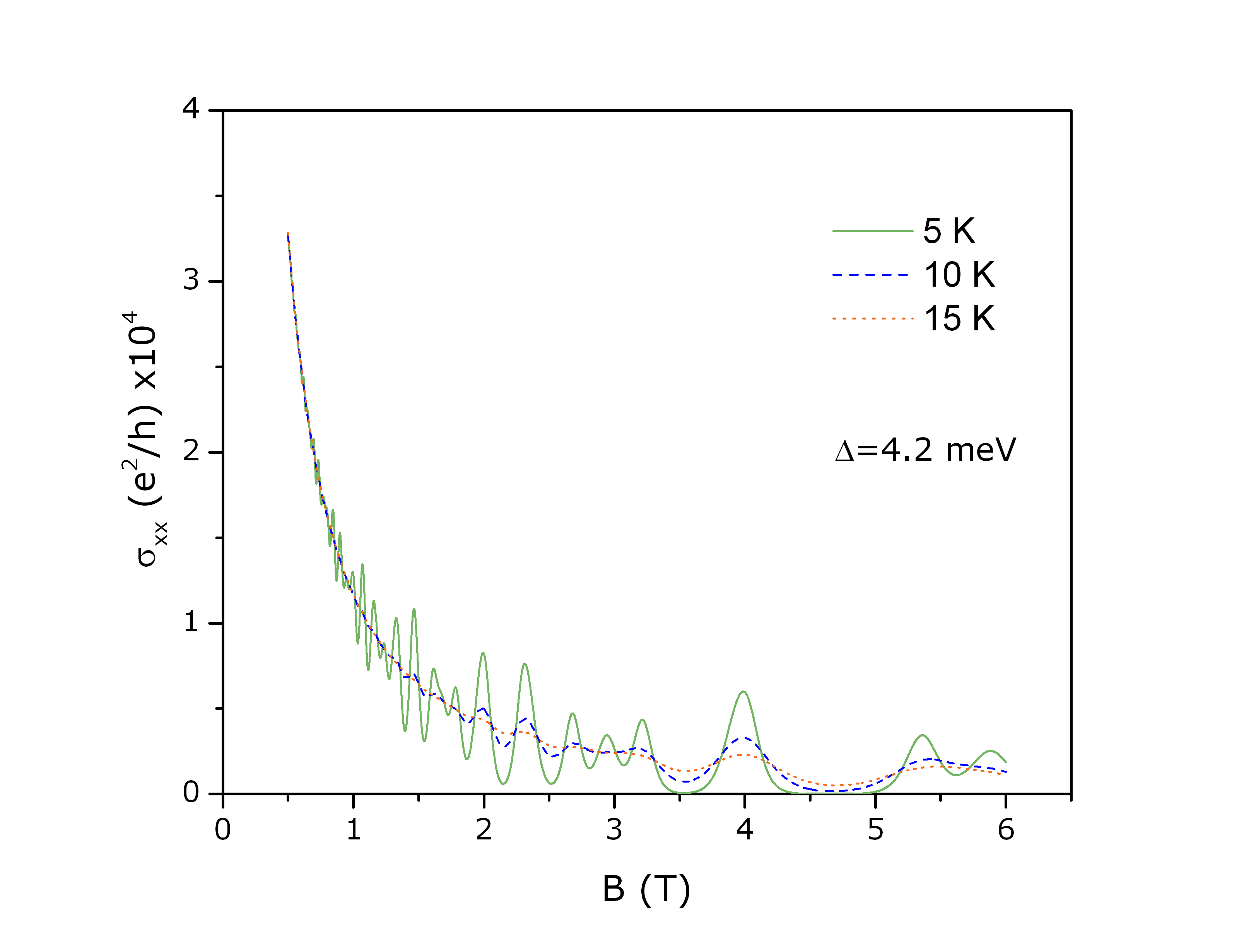}}}\hspace{0.5cm}
\subfigure[]{\includegraphics[width=8cm]{{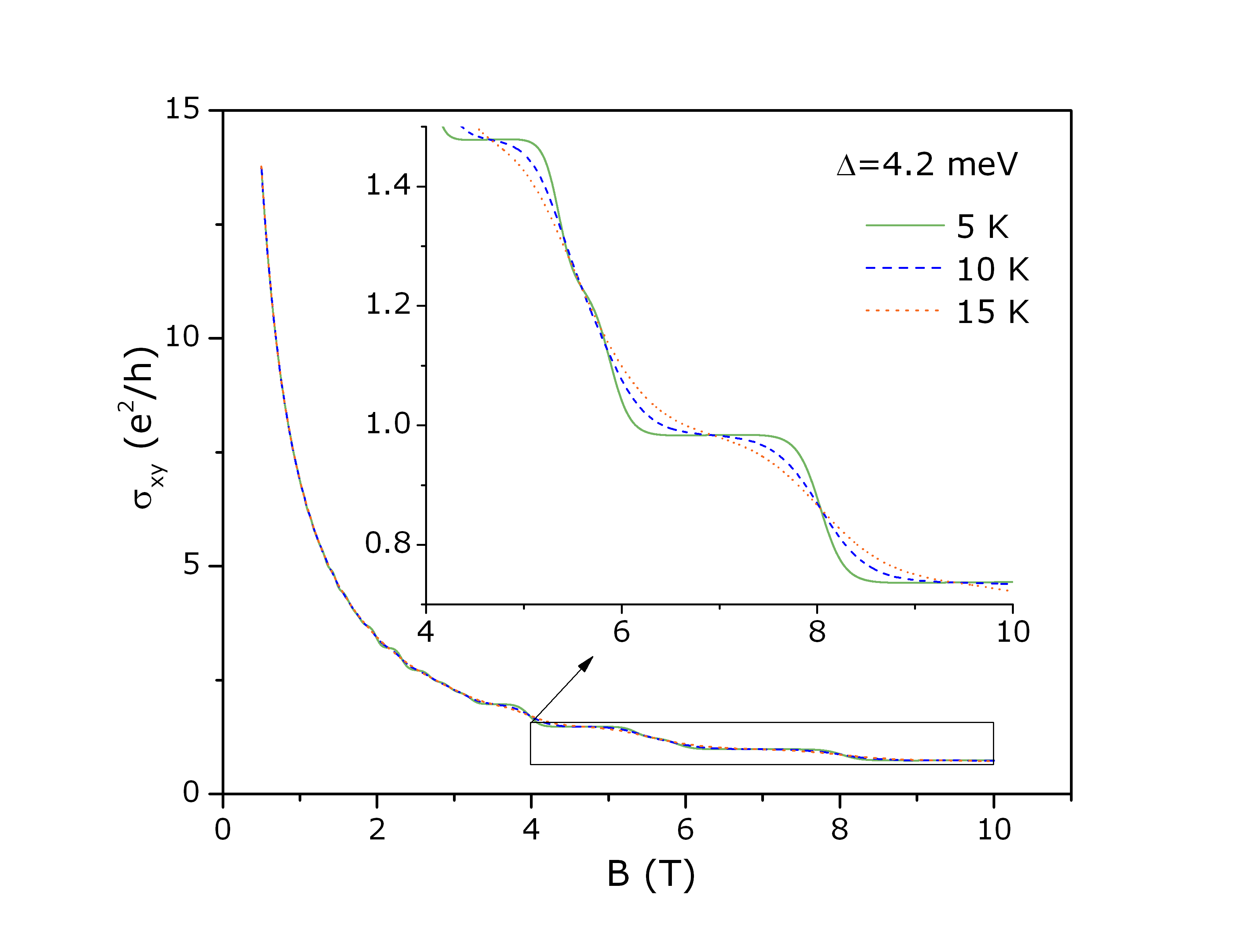}}}
\end{center}
\caption{(a) Collisional $\sigma_{xx}$ and (b) Hall $\sigma_{xy}$ conductivities as functions of chemical potential $\mu$ for $T=5$ K, $10$ K and $15$ K. }
\label{conductancefield}
\end{figure}

Figure~\ref{conductancefield} shows that $\sigma_{xx}$ and $\sigma_{xy}$ behave strikingly different as a function of magnetic field $B$. The former presents SdH oscillations with a well-resolved beating pattern, but the latter presents Hall plateaus with integer filling factors. This difference directly reflects the strong anisotropy of the magneto-conductivity in TIs. However, these have a common feature, being clear at strong magnetic field and low temperature. This is because the strong magnetic field is able to excite the Landau quantization, increasing the number of Landau states in each level and the spacing between two adjacent LLs. This leads to the pronounced splitting of the SdH peaks, and the well quantized Hall plateaus that decrease (in unit of $e^2/h$) one by one with increasing magnetic field. We stress that the SdH and Hall effects are quenched at low magnetic fields due to the tiny spacing between LLs. Moreover, the SdH amplitude increases with the magnetic field but decreases with temperature, since it is proportional to $\sqrt{B}$ and $1/T$ \cite{fu2007,naveed2020}.

\section{Seebeck and Nernst effect}

In this section, we study in details the thermopower $S$. 
First, we derive $S$ using Eq.~(\ref{thermopower}) and the conductivities in Eqs.~(\ref{sigmaxx}) and (\ref{sigmaxy}). 
Then, we discuss the general aspects of the numerical results obtained using the derived expressions.  

The 2D conductivity tensor in Eq.~(\ref{sigma}) leads to a 2D thermopower tensor in Eq.~(\ref{thermopower}), that can be cast as
\begin{align}
S=\left(
\begin{array}{cc}
S_{xx} & S_{xy} \\
S_{yx} & S_{yy}%
\end{array}%
\right),
\end{align}
where
\begin{align}
\label{sxx}
S_{xx} &= S_{yy}=\frac{\sigma _{xx}\kappa _{xx}+\sigma _{yx}\kappa _{yx}}{%
\sigma _{xx}^{2} + \sigma _{yx}^{2}}, \\
\label{sxy}
S_{yx} &=-S_{xy}=\frac{\sigma _{xx}\kappa _{yx}-\sigma _{yx}\kappa _{xx}}{%
\sigma _{xx}^{2} + \sigma _{yx}^{2}},
\end{align}
and
\begin{eqnarray}
\label{motttt1}
\kappa _{xx} &=&\kappa _{yy}=-\frac{\pi ^{2}k_{B}^{2}T}{3e}\frac{\partial
\sigma _{xx}}{\partial \mu }, \\
\label{motttt2}
\kappa _{yx} &=&-\kappa _{xy}=-\frac{\pi ^{2}k_{B}^{2}T}{3e}\frac{\partial
\sigma _{yx}}{\partial \mu }.
\end{eqnarray}
Thus, one can identify the Seebeck and Nernst coefficients as $S_{xx}$ in Eq.~(\ref{sxx}) and $S_{xy}$ in Eq.~(\ref{sxy}), respectively.

In order to calculate $S_{xx}$ and $S_{xy}$ we take the derivative of the conductivities in Eqs.~(\ref{sigmaxx}) and (\ref{sigmaxy}) yielding
\begin{widetext}
\begin{align}
\label{dsigmaxxdmu}
\frac{\partial \sigma _{xx}}{\partial \mu} &=\frac{e^{2}}{h}\frac{  N_{I} e^{4}}{8\pi^{2}\epsilon^{2}l_{c}^{2}k_{s}^{2}\hbar \omega_{c}k_B^2T^2} \times \nonumber \\ &\times \sum\limits_{n,\tau_{z},\lambda}\left[\left(2n+1\right) \cos^{4}\left(\frac{\theta}{2}\right) + (2n-1)\sin^{4}\left(\frac{\theta}{2}\right)-2n
\cos^{2}\left(\frac{\theta}{2}\right)\sin^{2}\left(\frac{\theta}{2}\right)\right]{\text{sech}^{2}\left(\xi^{\tau_z}_{n,\lambda}\right)\text{tanh}\left(\xi^{\tau_z}_{n,\lambda}\right)}, \\
\label{dsigmaxydmu}
\frac{\partial \sigma _{xy}}{\partial \mu} &=\frac{e^{2}}{h}\frac{1}{2k_BT}\sum\limits_{n,\tau_{z},\lambda}\sin^{2}\left(\theta\right)\left( n+\frac{1}{2}%
\right) \left[\text{sech}^{2}\left(\xi_{n+1,\lambda}^{\tau_{z}}\right) -\text{sech}^{2}\left(\xi_{n,\lambda}^{\tau_{z}}\right) \right].
\end{align}
\end{widetext}

In the regime where the collisional conductivity is much larger than the Hall conductivity, that is, 
$\left|\sigma_{xx}\right| \gg \left|\sigma_{xy}\right|$,  $S_{xx}$ and $S_{xy}$ assume simpler expressions, namely,
\begin{align}
\label{sxxapprox}
     S_{xx} \approx \frac{\kappa_{xx}}{\sigma_{xx}}=-\frac{\pi^{2}k_{B}^{2}}{3e }\frac{T}{\sigma_{xx}}\frac{\partial \sigma_{xx}}{\partial \mu}.
\end{align}
and
\begin{align}
\label{sxyapprox}
     S_{xy} \approx \frac{\kappa_{xy}}{\sigma_{xx}}=-\frac{\pi^{2}k_{B}^{2}}{3e }\frac{T}{\sigma_{xx}}\frac{\partial \sigma_{xy}}{\partial \mu}.
\end{align}

\begin{figure}
\begin{center}
\includegraphics[width=9cm]{{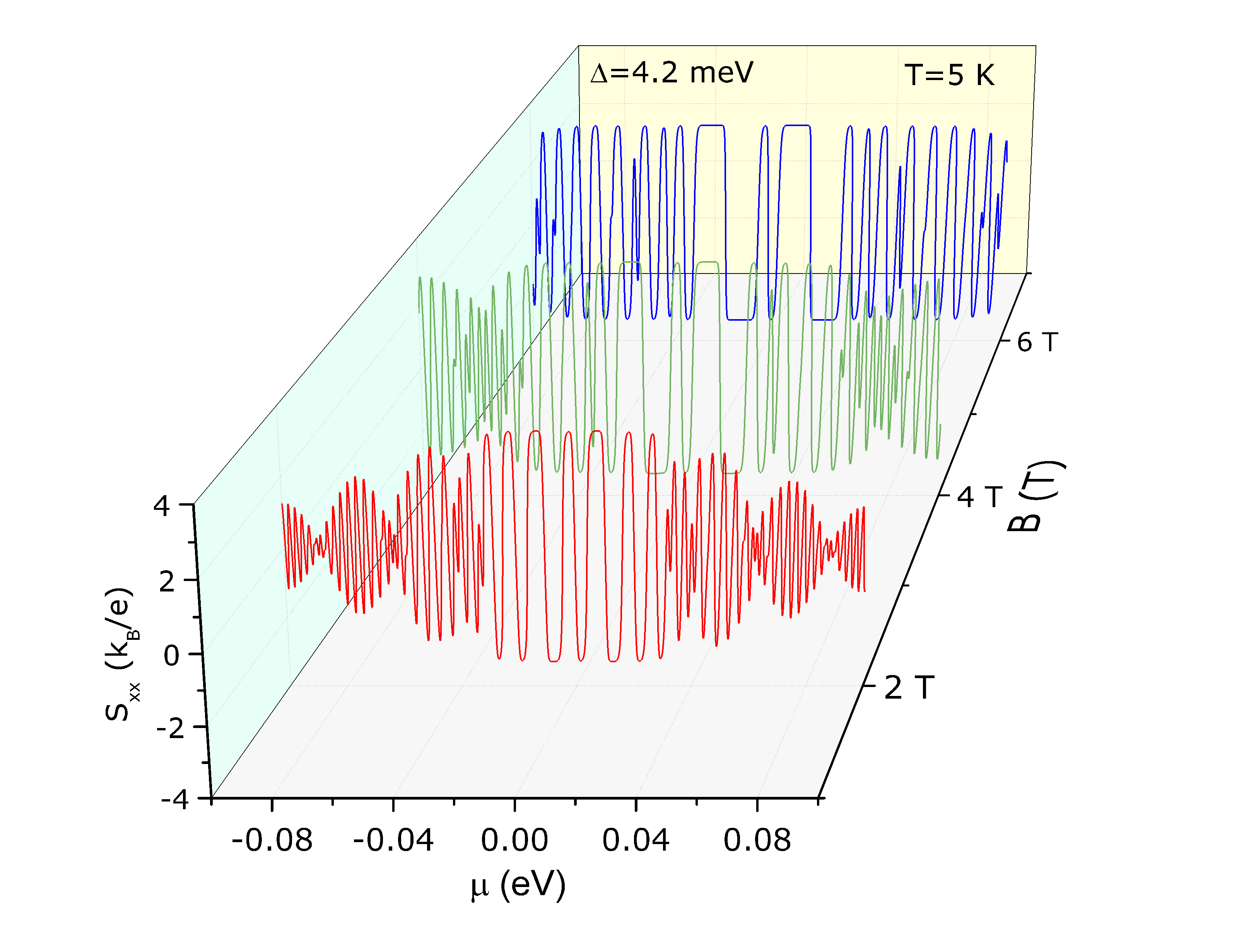}}
\end{center}
\caption{Seebeck coefficient $S_{xx}$ as a function of the chemical potential $\mu$ for some values of external magnetic field.}
\label{figsxxmub}
\end{figure}

Figure~\ref{figsxxmub} shows the behavior of the Seebeck coefficient $S_{xx}$ as a function of the chemical potential at $T=5$ K, for magnetic field values of $2$ T, $4$ T and $6$ T. 
Let us first analyze the behavior of $S_{xx}$ near the CNP ($\mu=0$) for $B=2$ T.
We find that $S_{xx}$ vanishes when the chemical potential $\mu$ crosses a LL, being negative for $\mu>E_{n,\lambda}^\tau$ and positive for $\mu<E_{n,\lambda}^\tau$.
The reason is that $S_{xx}$ captures the difference between the contribution from carriers with energies above and below $\mu$ inside the energy window of roughly 4 $k_BT$ around $\mu$.
Since the density of states around a LL is quite symmetric, when $\mu$ is on top of a LL, both types of carriers give nearly equal contributions that cancel out each other.
When $\mu$ is slightly above (below) a LL energy, this balance is broken and the negative (positive) contribution dominates.
In a nut shell, a local maximum (minimum) in $\sigma_{xx}(\mu)$ results in a positive (negative) slope in $S_{xx}(\mu)$.

Notice that, as we increase the magnetic field $B$ from $2$ T up to $6$ T, $S_{xx}$ does not experience significant changes around the CNP, having a typical linear dependence with $\mu$ \cite{gusynin2015spin}.
The magnetic field does not change the $0$-th LL's energies, given by Eq.~(\ref{energyspectrum}), that are responsible for the electronic transport around the CNP.  
Thus no energy shift is expected around the CNP as $B$ increases.
Since the intensity of $S_{xx}$ can be approximated by Eq.~(\ref{sxxapprox}) near those $0$-th LL's and the dependence of $\sigma_{xx}$ with $B$ lies in the pre-factor of Eq.~(\ref{sigmaxx}), we find that $S_{xx}$ does not depend on $B$ around the CNP, which is consistent with the data in Fig.~\ref{figsxxmub}.
On the other hand, the magnetic field shifts the higher energy LL's located far from the CNP. It creates the gaps where $S_{xx}$ achieves maximum intensity, as we can clearly see in Fig.~\ref{figsxxmub} for $B=6$ T. Thus, the magnetic field can be tuned to broaden the energy window where $S_{xx}$ reaches its maximum, making it robust against temperature fluctuations.

\begin{figure}
\begin{center}
\subfigure[]{\includegraphics[width=8cm]{{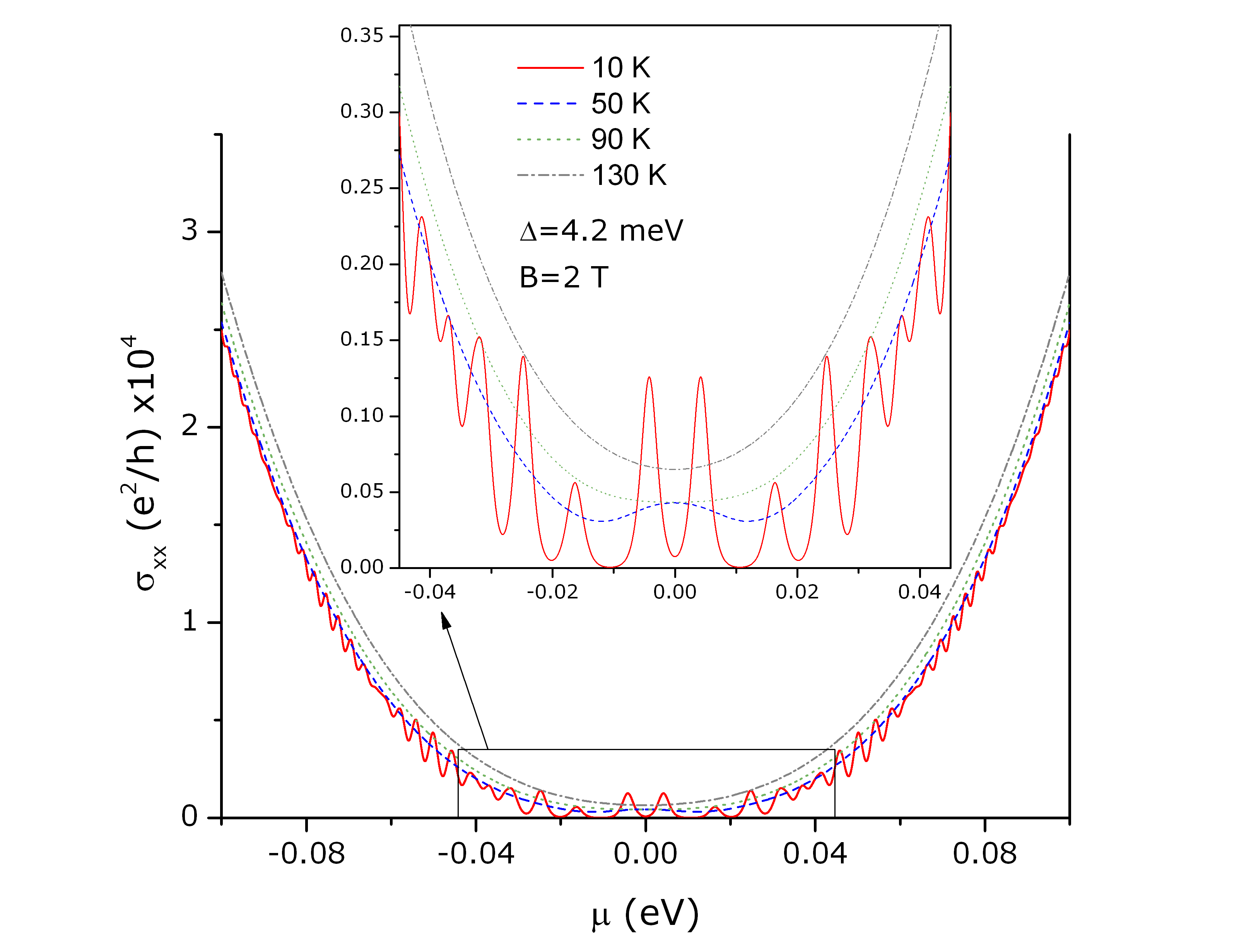}}}\hspace{0.5cm}
\subfigure[]{\includegraphics[width=8cm]{{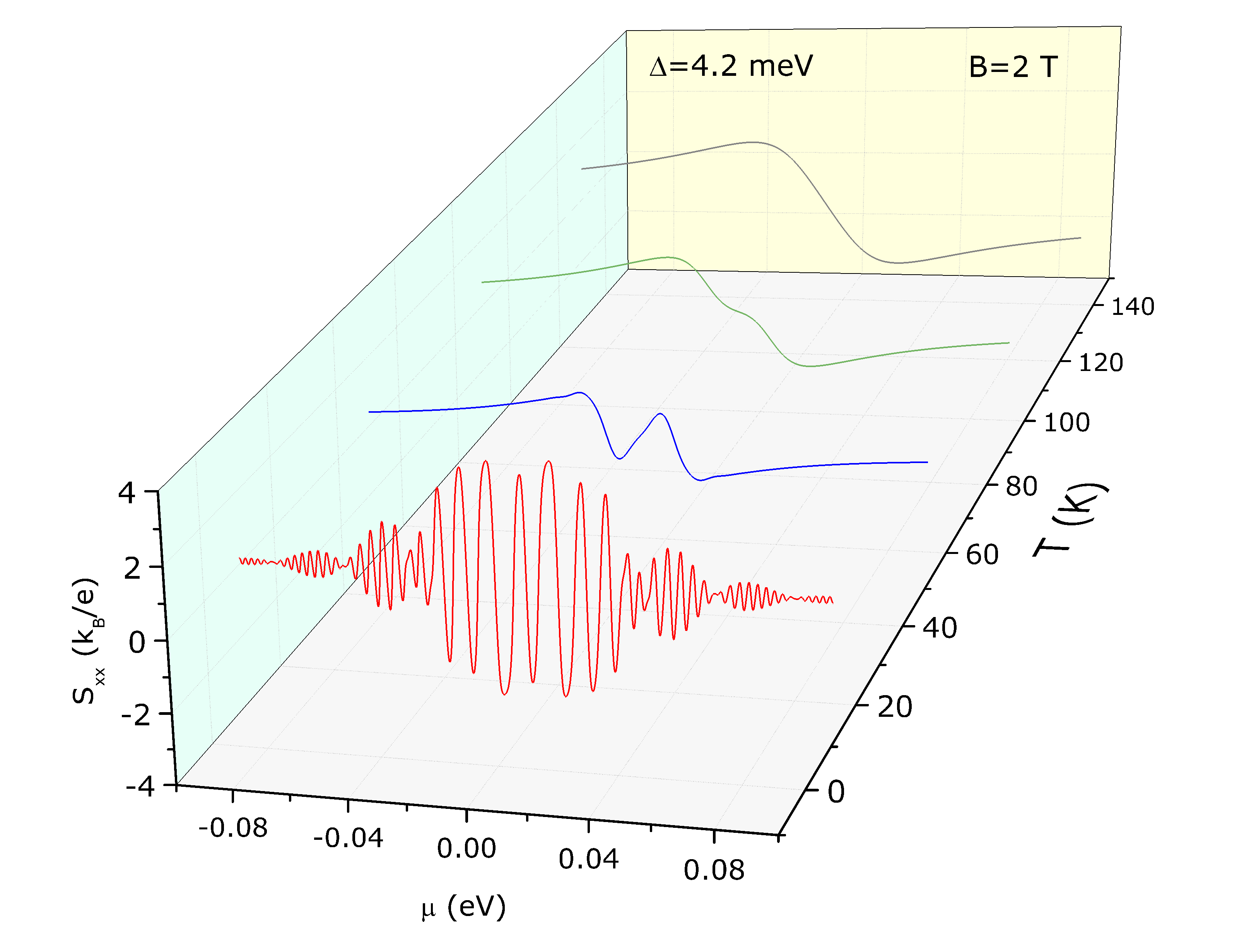}}}
\end{center}
\caption{(a) Collisional conductivity $\sigma_{xx}$ and (b) Seebeck coefficient $S_{xx}$ as functions of the chemical potential $\mu$ for some values of the temperature.}
\label{figsxxmut}
\end{figure}

For magnetic field values of $B=6$ T, as in Fig.~\ref{figsxxmub}, the beating pattern of $S_{xx}$ is destroyed. $S_{xx}$ increases up to the maximum absolute value reported for this case, i.e., 3 $k_B/e$. 
Since the separation between consecutive LL's increases with $B$, the overlap between the conducting states contributing to $S_{xx}$ diminishes as $B$ increases. This effect smashes down the beating pattern as the LL's tend to contribute individually, similarly to the $0$-th LL's around the CNP, without interference from other neighbor LL. 
Additionally, intensifying $B$ causes an increasing in the occupied states in each LL, yielding an increasing in the intensity of $S_{xx}$.

Fig.~\ref{figsxxmut} presents the Seebeck coefficient $S_{xx}$ and the collisional conductivity $\sigma_{xx}$ as functions of the chemical potential $\mu$ for different temperatures values, ranging from $T=10$ K to $T=130$ K, at $B=$ 2 T. 
We note that, at low temperatures, the conductivity $\sigma_{xx}$ reaches high values around the CNP at the $0$-th LL's, as discussed in the previous paragraphs. However, as the temperature increases, there is a general increasing of $\sigma_{xx}$ in the whole chemical potential range. The inset of Fig.~\ref{figsxxmut}(a) shows the region around the CNP in detail. The temperature smashes down the beating pattern due to strain. As a consequence, the Seebeck coefficient $S_{xx}$ curve, shown in Fig.~\ref{figsxxmut}(b), becomes smoother increasing temperature. 

Thermal fluctuations mix neighbor states as the temperature increases, and suppress the electronic gaps between consecutive LL's. This suppression depends on the ratio between the energy separation between LL's, that is ruled by $\hbar\omega_c$, and the thermal broadening of the states around $\mu$, roughly of the order of $k_BT$. 
The maximum value of $S_{xx}$, as shown in Fig.~\ref{figsxxmut}(b), is equal to $3.2\ k_B/e$, $1.2\ k_B/e$, $1.6\ k_B/e$ and $1.9\ k_B/e$ for the temperatures $T=10$ K, $T=50$ K, $T=90$ K and $T=130$ K respectively, indicating that it changes non-monotonically with temperature. 
Therefore, one needs to control both the temperature and the chemical potential to tune the maximum value of $S_{xx}$.

The Seebeck coefficient $S_{xx}$ in Fig.~\ref{figsxxmut}(b) presents a change in the slope around the CNP as the temperature increases.
As we discussed at Fig.~\ref{figsxxmub}, this is a consequence of the change in the slope of the curve $\sigma_{xx}(\mu)$ around the CNP.
The conduction profile shown in the inset of Fig.~\ref{figsxxmut}(a) exhibit a transition from a minimum to a maximum at $\mu=0$ when the temperature changes from $T=10$ K to $T=50$ K, respectively.
This transition is a consequence of the broadening and overlap between the $0$-th LL's around the CNP. 
But as $T$ increases even more to $90$ K or $130$ K, $\sigma_{xx}(\mu)$ becomes very smooth, rendering the minimum conductivity at the CNP shown in Fig.~\ref{figsxxmut}(a) and the the negative slope of $S_{xx}(\mu)$ around the CNP shown in Fig.~\ref{figsxxmut}(b).

\begin{figure}
\begin{center}
\subfigure[]{\includegraphics[width=8cm]{{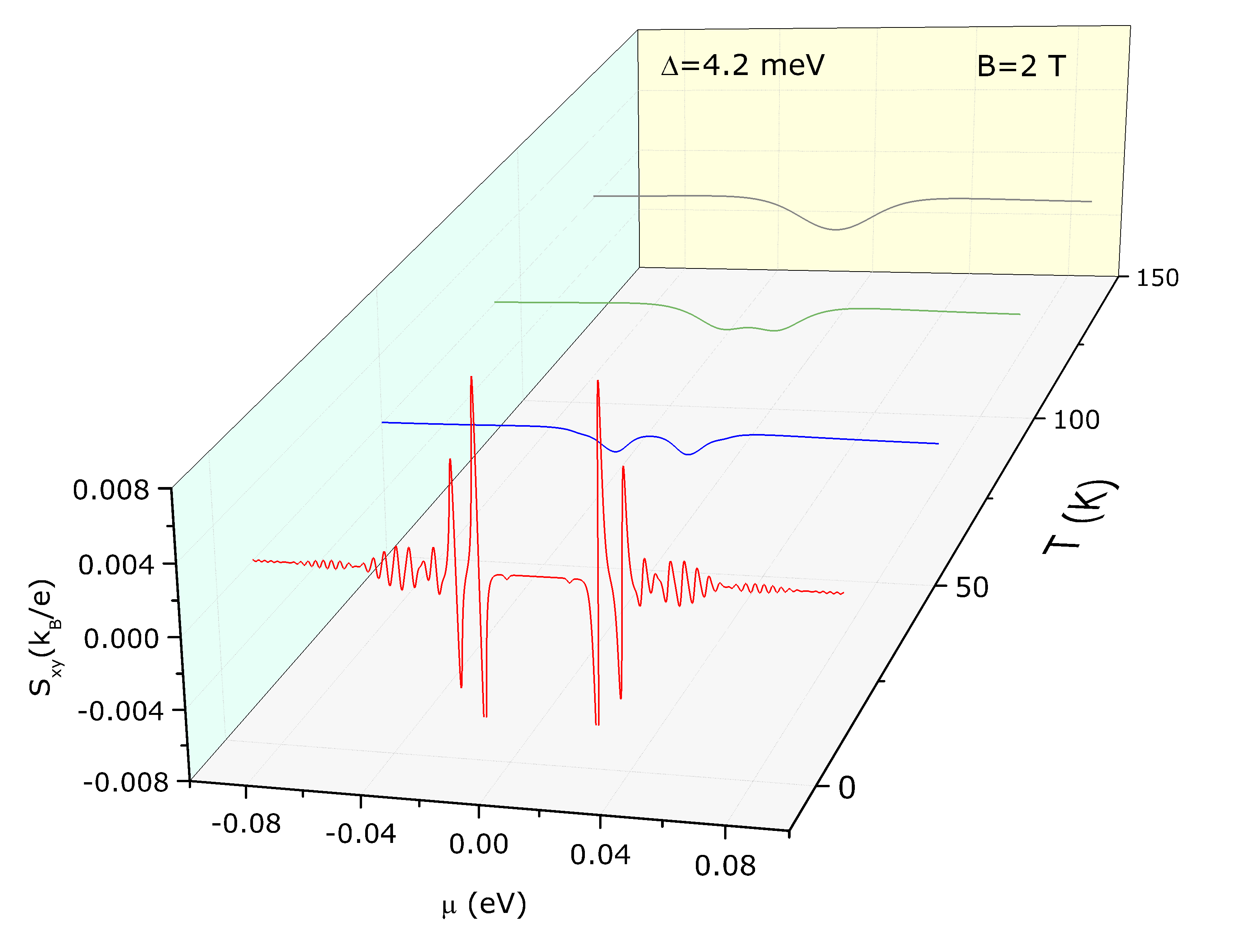}}}\hspace{0.5cm}
\subfigure[]{\includegraphics[width=8cm]{{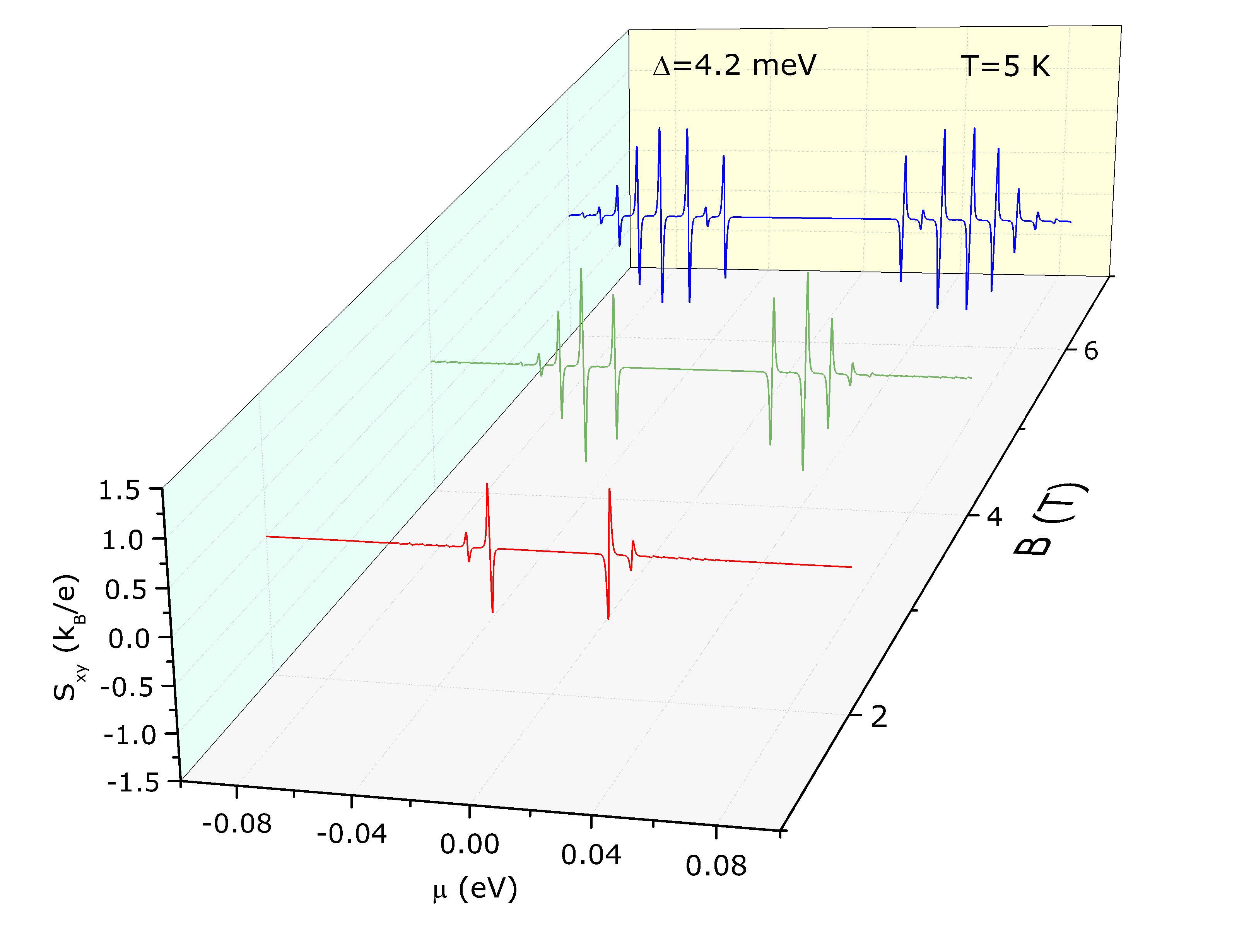}}}
\end{center}
\caption{
Nernst coefficient $S_{xy}$ as a function of the chemical potential $\mu$. In the panel (a) the temperature is changed for a constant magnetic field of $B=2$ T; while in panel (b) the magnetic field is changed for a constant temperature of $5$ K.}
\label{figsxymutb}
\end{figure}

Now we turn our attention to the Nernst effect, with help of Fig.~\ref{figsxymutb}. We present the Nernst coefficient $S_{xy}$ as a function of chemical potential $\mu$ for different temperatures in Fig.~\ref{figsxymutb}(a) and for different magnetic fields in Fig.~\ref{figsxymutb}(b). Figure~\ref{figsxymutb}(a) shows that $S_{xy}$ also presents a beating pattern due to strain, similar to the one seen in $S_{xx}$ at Fig.~\ref{figsxxmut}(b). As discussed before, the thermal effects also smooth the curve as the temperature increases from $T=10$ K to $T=130$ K.
On the other hand, the parity of the curves are not similar as $S_{xy}$ presents an even parity with $\mu$, while $S_{xx}$ presents an odd parity. 
This is a direct consequence of the opposite parities of the transverse conductivity $\sigma_{xy}$ and the longitudinal conductivity $\sigma_{xx}$, as seen in Fig.~\ref{figsigmamub}.
One can understand this behavior by noticing Fig.~\ref{figsigmamub}(b). For $\mu>0$ the number of transverse propagating states above $\mu$ is greater than the number of transverse propagating states below $\mu$, while for $\mu<0$ the number of available states below $\mu$ is greater than the one above $\mu$. 
However, $\sigma_{xy}$ is negative for $\mu<0$, indicating propagation in the opposite direction with respect to the states with $\mu>0$. The difference between the number of propagating states above and below $\mu$ for $\mu>0$ and $\mu<0$ is compensated by the change in propagation direction. Thus, the response in $S_{xy}$ has the same sign in both cases. 

We find that, in general, the Nernst effect is stronger at chemical potentials matching the LL's energies given by Eq.~(\ref{energyspectrum}). 
For $\mu$ at any LL energy, that is, at the transition between two consecutive plateaus of $\sigma_{xy}$ in Fig.~\ref{figsigmamub}(b), the difference on the number of propagating modes above and below $\mu$ reaches its absolute maximum value, enhancing the Nernst effect. Thus, away from any LL, in a chemical potential window where the number of available transverse channels is the same above and below $\mu$, the Nernst effect, quantified by $S_{xy}$, vanishes. This physical picture explains the behavior we see in Fig.~\ref{figsxymutb}(b), where $S_{xy}$ peaks at a plateau transitions of $\sigma_{xy}$ and the number of peaks increases with the magnetic field, since higher $B$ values promote plateau transitions with better definitions.
Moreover, the value of $S_{xy}$ is zero around the CNP and the non-topological gap in $S_{xy}$ increases with $B$ as the plateau transitions in $\sigma_{xy}$ move far away from the CNP as $B$ increases [see Fig.~\ref{figsigmamub}(b)].

\begin{figure}
\begin{center}
\subfigure[]{\includegraphics[width=8cm]{{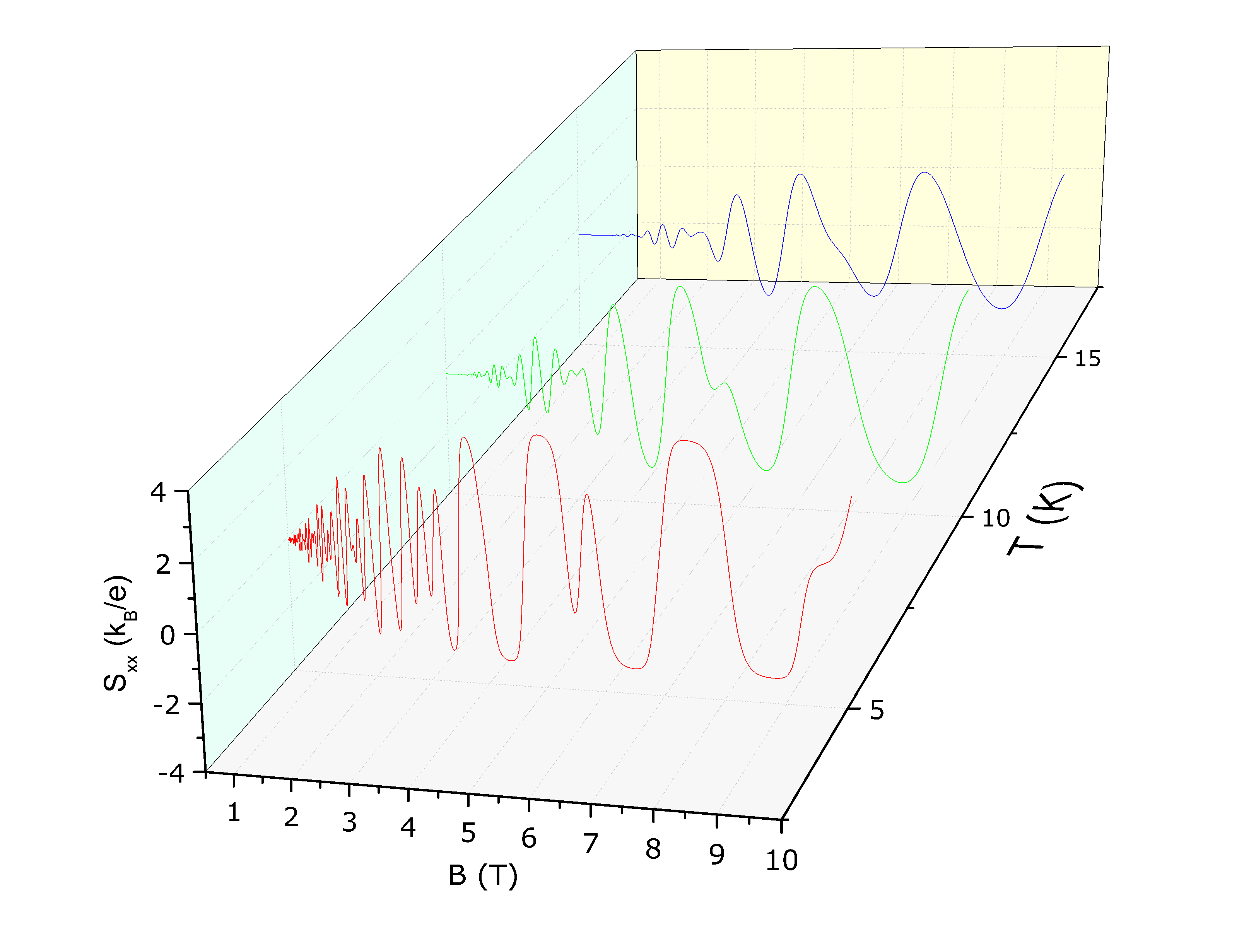}}}\hspace{0.5cm}
\subfigure[]{\includegraphics[width=8cm]{{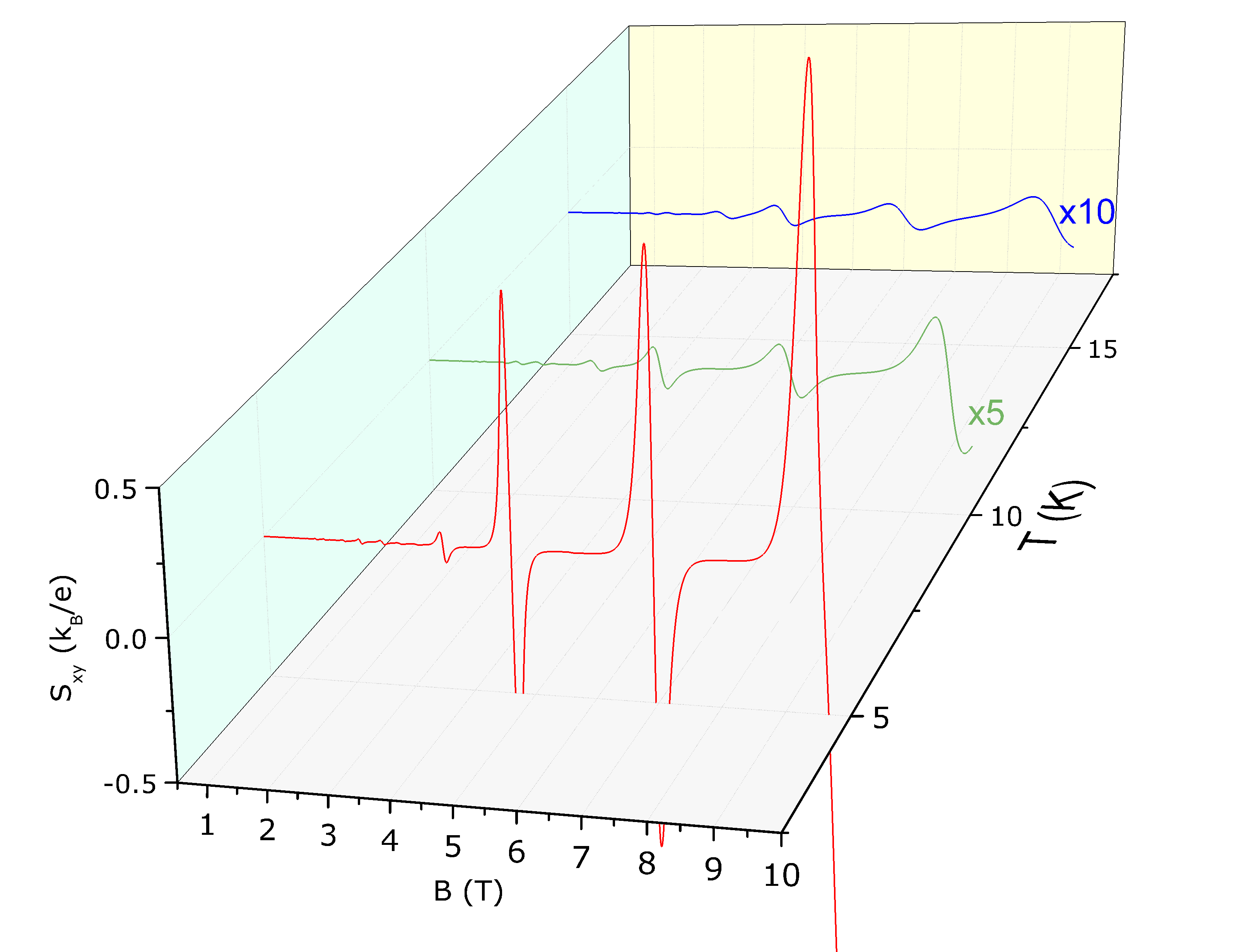}}}
\end{center}
\caption{(a) Seebeck and (b) Nernst coefficients as functions of the external magnetic field $B$ for different values of the temperature. The chemical potential is $\mu=54$ meV and the strain energy is $\Delta=4.2$ meV.}
\label{figsbt}
\end{figure}

Finally, we display the dependence of the Seebeck $S_{xx}$ and Nernst $S_{xy}$ coefficients with respect to the magnetic field $B$ in Figs.~\ref{figsbt}(a) and \ref{figsbt}(b), respectively, for different temperatures.
As expected, both $S_{xx}$ and $S_{xy}$ show SdH oscillations with the strain-induced beating pattern that fades away with increasing $T$, as we first discussed in Fig.~\ref{figsigmamub}. Additionally, their oscillation amplitudes tends to increase with $B$ since for a constant chemical potential $\mu$ the number of filled LL's decreases with $B$ as they cross $\mu$ and depopulate as $B$ increases.
Although both $S_{xx}$ and $S_{xy}$ oscillate, a further analysis (not shown here) shows that they are out of phase by $\pi$. 

Notice that the magnitude of the Nernst coefficient $S_{xy}$ reaches values in the order of $k_{B}/e \sim 86$ $\mu$V/K, which is in the same order of magnitude of the one found in bulk graphene \cite{duan2018}, graphene nanoribbons \cite{hozana2019} and in other two-dimensional Dirac materials such as phosphorene, stanene, and germanene \cite{gusynin2015spin}. In normal non-magnetic metals the Nernst signal is negligibly small ($\sim 10^{-2}$ $\mu$V/K) \cite{gusynin2015spin}. We emphasize that in some Dirac materials the chemical potential $\mu$ can be tuned, which opens up the possibility to experimentally set the regime that maximizes both Seebeck and Nernst effects. 

\section{Conclusions}

In this work, we have explored the Seebeck and Nernst effect in the case of a strained HgTe material subjected to a constant and uniform magnetic field perpendicular to it. The approximation we have taken to describe its energy levels corresponds to the 2D non-ideal Dirac quasi-particle Hamiltonian. 

For the calculation of the electrical conductivity tensor, it is obtained analytically through the Kubo-Greenwood formalism previously discussed in Ref. \cite{tahir2012quantum,tahir2016}. The collisional conductivity shows a well-resolved beating pattern of SdH oscillation away from CNP, indicating that these oscillations from different surfaces are out of phase. On the other hand, the Hall conductivity shows some extra plateaus due to the strain imposed by the CdTe substrate on the HgTe. Both conductivities have a common feature regarding the effect of the strain on them; it is more robust (evident) at the high-magnetic field and low temperature. Additionally, it is obtained that the magnitude of the collisional conductivity is much larger than the Hall conductivity, and therefore it will be the collisional conductivity that will mainly interfere in the results reported for the Seebeck coefficient. Consistent with the conductivity tensor results, the Seebeck and Nernst thermoelectric coefficients show oscillatory behavior due to SdH oscillations. The pattern of the oscillations of the thermoelectric quantities is strongly modulated by the intensity of the strain on HgTe imposed by the substrate. The Nernst coefficient found has even parity, while the Seebeck coefficient is odd in coherence with the opposite parities of the Hall conductivity (or transverse conductivity) and the collisional conductivity (or longitudinal conductivity)  presented in this work. We report values for the Seebeck close to 3.2 $k_{B}/e$ and the Nernst effect of 1.2 $k_{B}/e$ in the energy regimes of the strain order.


\section{Acknowledgments}

F.J.P. acknowledges support from ANID Fondecyt, Iniciaci\'on en Investigaci\'on 2020 grant No. 11200032, and the financial support of USM-DGIIE.
L.R.F.L. acknowledges financial support
from the Brazilian Institute of Science and Technology
(INCT) in Carbon Nanomaterials and the Brazilian
agency FAPERJ. P.V. acknowledges support from ANID Fondecyt grant No. 1210312, O.N. to  ANID PIA/Basal grant No.  AFB18000.

\bibliography{article_bibfile}

\end{document}